\documentclass[aip, jcp, amsmath, amssymb, reprint, longbibliography]{revtex4-1}

\usepackage{graphicx}
\usepackage{dcolumn}
\usepackage{bm}
\usepackage{color}

\begin{document}


\title{Self-assembly and glass-formation in a lattice model of telechelic polymer melts: Influence of stiffness of the sticky bonds}

\author{Wen-Sheng Xu}
\email{wsxu@uchicago.edu}
\affiliation{James Franck Institute, The University of Chicago, Chicago, Illinois 60637, USA}

\author{Karl F. Freed}
\email{freed@uchicago.edu}
\affiliation{James Franck Institute, The University of Chicago, Chicago, Illinois 60637, USA}
\affiliation{Department of Chemistry, The University of Chicago, Chicago, Illinois 60637, USA}

\date{\today}

\begin{abstract}
Telechelic polymers are chain macromolecules that may self-assemble through the association of their two mono-functional end groups (called ``stickers''). A deep understanding of the relation between microscopic molecular details and the macroscopic physical properties of telechelic polymers is important in guiding the rational design of telechelic polymer materials with desired properties. The lattice cluster theory (LCT) for strongly interacting, self-assembling telechelic polymers provides a theoretical tool that enables establishing the connections between important microscopic molecular details of self-assembling polymers and their bulk thermodynamics. The original LCT for self-assembly of telechelic polymers considers a model of fully flexible linear chains [J. Dudowicz and K. F. Freed, J. Chem. Phys. \textbf{136}, 064902 (2012)], while our recent work introduces a significant improvement to the LCT by including a description of chain semiflexibility for the bonds within each individual telecheic chain [W.-S. Xu and K. F. Freed, J. Chem. Phys. \textbf{143}, 024901 (2015)], but the physically associative (or called ``sticky'') bonds between the ends of the telechelics are left as fully flexible. Motivated by the ubiquitous presence of steric constraints on the association of real telechelic polymers that impart an additional degree of bond stiffness (or rigidity), the present paper further extends the LCT to permit the sticky bonds to be semiflexible but to have a stiffness differing from that within each telechelic chain. An analytical expression for the Helmholtz free energy is provided for this model of linear telechelic polymer melts, and illustrative calculations demonstrate the significant influence of the stiffness of the sticky bonds on the self-assembly and thermodynamics of telechelic polymers. A brief discussion is also provided for the impact of self-assembly on glass-formation by combining the LCT description for this extended model of telechelic polymers with the Adam-Gibbs relation between the structural relaxation time and the configurational entropy.
\end{abstract}



\maketitle

\section{Introduction}

Telechelic polymers provide a striking example of associating macromolecules that are capable of supramolecular self-assembly.~\cite{Polymer_49_1425} The distinctive properties of telechelic polymers arise from their mono-functional end groups (called ``stickers'') that permit the reversible formation and breakage of physical bonds during the dynamical self-assembly, thereby opening the prospect of many new applications~\cite{POC_7_289, Book_Goodman, Polymer_45_3527, Nature_453_171} that are generally inaccessible by conventional methods of polymerization. While the increasing scientific interest in telechelic polymers and their technological importance have motivated a number of theoretical~\cite{Mac_28_1066, Mac_28_7879, JCP_110_1781, Mac_33_1425, Mac_33_1443, JCP_119_6916, Lan_20_7860, JPSB_45_3285, JCP_131_144906, JPCB_114_12298} and numerical~\cite{Mac_20_1999, JCP_110_6039, EPL_59_384, Polymer_45_3961, JPSB_43_796, PRL_96_187802, PRL_109_238301, JCP_126_044907, JPCM_20_335103, Mac_47_4118, Mac_47_6946, JCP_143_243117, SM_2016} investigations of their physical behavior, a substantial challenge confronts the development of analytical theories for the connection between microscopic monomer details and the nature of the self-assembly and thermodynamics.

The lattice cluster theory (LCT)~\cite{JCP_87_7272, Mac_24_5076, ACP_103_335, APS_183_63} describes the thermodynamics of polymer systems by employing an intermediate level of coarse-grained models that retains the essential features of molecular structure and interactions in polymer fluids and that enables investigating the impact of various molecular characteristics upon the thermodynamic properties of polymer systems. The extension of the LCT developed here considers inclusion of strong interactions between the stickers in telechelic polymers, a treatment that poses the need to reformulate the LCT.~\cite{JCP_130_061103, JCP_136_064902} The initial studies by Dudowicz and Freed~\cite{JCP_136_064902} consider, for simplicity, models of fully flexible linear telechelic polymers. Hence, several improvements are desirable within the LCT for telechelic polymers. For instance, our recent work~\cite{JCP_143_024901, JCP_143_024902} begins to address the role of chain semiflexibility in determining the thermodynamic properties of telechelic polymers. Following the original treatment,~\cite{ACP_103_335} chain semiflexibility is described in our previous work~\cite{JCP_143_024901} by introducing a bending energy penalty whenever a pair of consecutive bonds from the \textit{same} chain lies along orthogonal directions. This description implies that the physical bonds between the stickers are fully flexible. Nevertheless, the physically sticky bonds in real telechelic polymers must possess a degree of bond stiffness (or rigidity) due to steric interactions of the stickers. For example, the formation of N-H-O hydrogen bonds is restricted to occur over a narrow range of angles. Therefore, a theory for the influence of the stiffness of sticky bonds on the self-assembly and thermodynamics remains to be developed. The present paper further extends the LCT for linear telechelic polymers by introducing a separate bending energy penalty to a pair of sequential orthogonal bonds containing one sticky bond, thereby permitting the sticky bonds to be semiflexible. The stiffness of the sticky bonds turns out to greatly influence the self-assembly and thermodynamics of telechelic polymers.

Section II provides a description of the LCT model for semiflexible linear telechelic polymers, along with a summary of the Helmholtz free energy. Section III begins by demonstrating the strong dependence of the stiffness of the sticky bonds on the average degree of self-assembly in telechelic polymers. Previous work~\cite{JCP_143_024902} for telechelic polymers with fully flexible sticky bonds indicates that the average degree of self-assembly is elevated by chain stiffness when either the polymer filling fraction $\phi$ or the temperature $T$ is high, but diminishes as the chains stiffen when both $\phi$ and $T$ are low. These general trends are shown to likewise occur in telechelic polymers with semiflexible sticky bonds. We further examine how the stiffness of the sticky bonds influences this behavior. Section III then illustrates the great influence of the stiffness of sticky bonds on the self-assembly transition. A brief discussion follows in Sec. III of glass-formation that emerges for self-assembling telechelic polymers by combining the current extension of the LCT with the Adam-Gibbs relation~\cite{JCP_43_139} [i.e., the resultant generalized entropy theory (GET)~\cite{ACP_137_125}] between the structural relaxation time and the configurational entropy.

\section{Lattice cluster theory for semiflexible linear telechelic polymer melts}

This section introduces the lattice model of semiflexible linear telechelic polymer melts considered in the present work, followed by a summary of the Helmholtz free energy derived for this model.

\subsection{Lattice model of semiflexible linear telechelic polymer melts}

The lattice model of polymers conventionally employs a $d$-dimensional hypercubic lattice with $N_l$ lattice sites, each with $z=2d$ nearest neighbors. The present work considers a compressible melt~\footnote{Note that the present paper discusses the model and results with reference to compressible melts. The mathematical equivalence between the excess thermodynamic properties of a compressible melt and an incompressible solution enables drawing conclusions for both types of systems. The model for an incompressible solution consists of solvent molecules, each of which occupy a single lattice site, replacing the empty lattice sites. The free energy expression for the compressible polymer melt is isomorphic to that for the incompressible polymer solution, with the microscopic cohesive interaction parameter $\epsilon$ being replaced by the exchange energy $\epsilon_{\text{ex}}=\epsilon_{pp}+\epsilon_{ss}-2\epsilon_{ps}$, where $\epsilon_{pp}$, $\epsilon_{ss}$, and $\epsilon_{ps}$ represent the strengths of the nearest neighbor interaction between two polymer segments, two solvent molecules, and a polymer segment and a solvent molecule, respectively.} consisting of $m$ linear chains, where the length of each chain is given by the number $M$ of united atom groups (also called ``beads'' or ``segments'' for simplicity) in a single chain. Since the system is compressible, each lattice site is either empty or occupied by a bead, thereby producing the filling fraction of the polymer segments as $\phi=mM/N_l$.

\begin{figure}[tb]
	\centering
	\includegraphics[angle=0,width=0.45\textwidth]{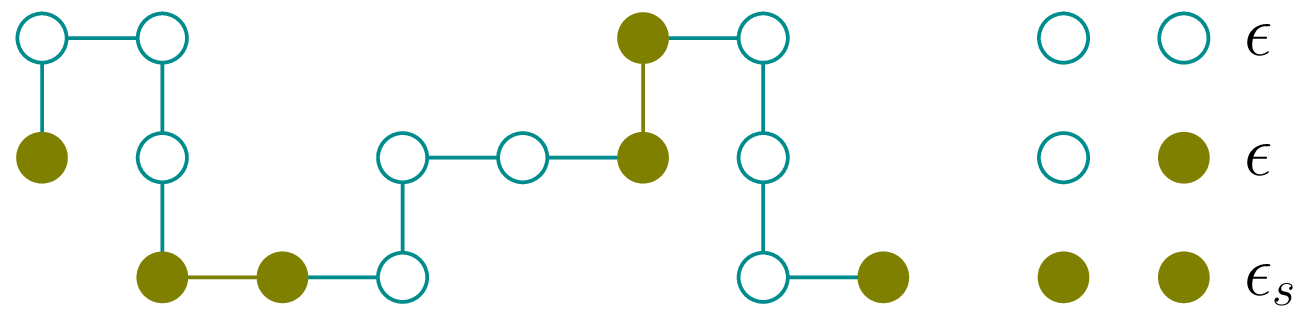}
	\caption{Illustration of the lattice model for a self-assembled linear cluster formed by three telechelic polymer chains, each with $M=5$ united atom groups. Solid circles (called stickers) designate the ends of the chains that can participate in strong sticky interactions, while open circles denote united atom groups in the chain interior. Lines linking two stickers denote the physically sticky bonds, while the other lines represent the chemical bonds between two consecutive united atom groups along the same chain. As shown in the figure, the model prescribes different nearest neighbor interaction energies $\epsilon$ and $\epsilon_s$ for ordinary and sticker-sticker interactions, respectively.}
\end{figure}

The lattice model accounts for the basic characteristics of telechelic polymers by first distinguishing the end segments of each chain (represented as solid circles in Fig. 1 and called stickers) from the other united atom groups lying in the chain interior (depicted by open circles in Fig. 1 and called non-stickers). As introduced in Ref.~\citenum{JCP_136_064902}, two stickers can form a physically sticky ``bond'' and interact with an enhanced attractive sticky interaction energy $\epsilon_s$ when they are located on nearest-neighbor lattice sites, thereby allowing the system to self-assemble upon cooling. Nearest-neighbor attractive interactions between two non-stickers as well as between a sticker and a non-sticker are described by the microscopic cohesive energy parameter $\epsilon$ (see Fig. 1). By convention, $\epsilon$ is treated as positive for attractive nearest neighbor interactions, while $\epsilon_s$ is defined as negative for attractive interactions. As in real telechelic polymers, the sticky interaction strength $|\epsilon_s|$ may greatly exceed the microscopic ordinary cohesive interaction strength $\epsilon$. The latter fact introduces the need for reformulating the LCT to treat polymer systems with both weak and strong interactions rather than just the high temperature series expansion inherent in the original LCT and inapplicable for strong interactions.~\cite{JCP_130_061103, JCP_136_064902} For simplicity, the model allows the stickers at each end of the telechelics to be mono-functional, implying that each sticker can only participate in one sticky interaction. In addition, the present model allows both cyclic and linear associative clusters to form upon cooling, in accord with previous work~\cite{ JCP_136_064902, JCP_136_244904} and the analysis of Jacobson and Stockmayer.~\cite{ JCP_18_1600}

\begin{figure}[tb]
	\centering
	\includegraphics[angle=0,width=0.2\textwidth]{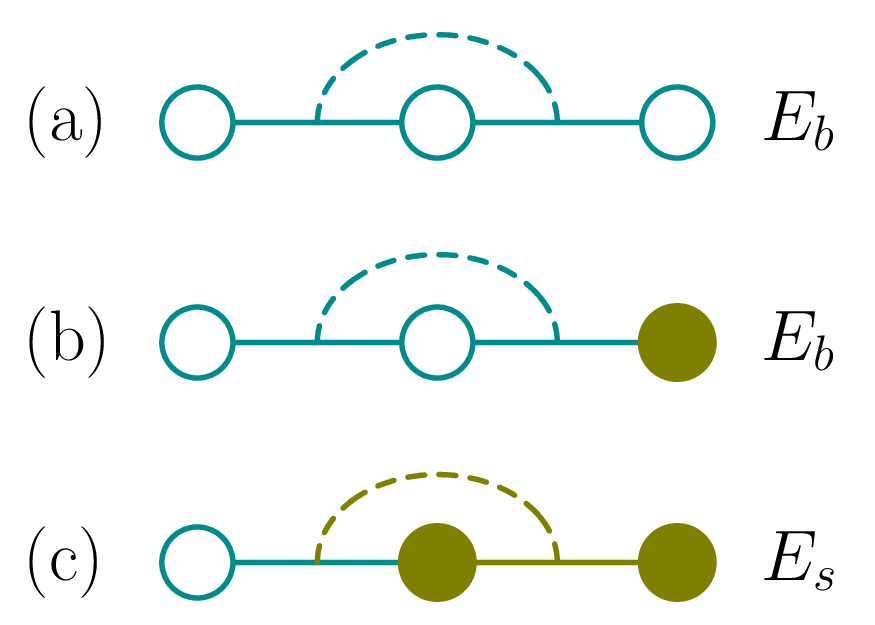}
	\caption{Illustration of including bending constraints in the lattice cluster theory for linear telechelic polymers. Bending constraints are depicted by the presence of dashed curved lines connecting pairs of consecutive bonds. While (a) and (b) illustrate examples where the ordinary bending rigidity parameter $E_b$ describes the stiffness of bonds within the same chain, (c) indicates that a separate sticky bending rigidity parameter $E_s$ is introduced to quantify the stiffness of a pair of bonds that includes one sticky bond. The figure exhibits the simplest diagrams consisting of two successive bonds as an illustration, but the same convention applies for all other diagrams.}
\end{figure}

Chain semiflexibility represents another important feature of real polymers and thus of telechelics. The LCT~\cite{ACP_103_335} traditionally incorporates chain semiflexibility following Flory~\cite{Flory_1956} by introducing a bending energy penalty $E_b$ (alternatively called the bending rigidity parameter) whenever a pair of consecutive bonds from a single chain lies along orthogonal directions. Our previous work~\cite{JCP_143_024901} for telechelic polymer melts adopts the same model for chain semiflexibility. Specifically, taking the diagram consisting of two successive bonds as an example, the previous theory~\cite{JCP_143_024901} considers only the two diagrams shown in Figs. 2(a) and 2(b) to describe bending constraints between pairs of bonds within a chain. This treatment, in turn, implies that the sticky bonds are fully flexible in the previous model.~\cite{JCP_143_024901} In order to more realistically represent the stiffness imparted by steric interactions to the sticky bonds in real telechelic polymers (e.g., bond angle constraints on hydrogen bonds), the present work introduces a separate bending rigidity parameter $E_s$ for each pair of sequential orthogonal bonds containing one sticky bond [see Fig. 2(c)]. For convenience, $E_b$ and $E_s$ are called the ordinary and sticky bending rigidity parameters, respectively. As shown in Sec. III, the thermodynamics and glass-formation of telechelic polymers are greatly influenced by the stiffness of the sticky bonds, as expected. We note that both $E_b$ and $E_s$ may be tuned in real telechelic polymers by altering the size and/or shape of the chemical groups, by introducing modifications to steric interactions hindering the development of sticky bonds, and/or by adjusting the polarity of the sticky units, features that are standard tools of synthetic chemists.

\subsection{Free energy of semiflexible linear telechelic polymer melts}

Since no new technical problems are posed by the addition of the sticky bending  constraints, we only summarize the results that are required for using the theory. References~\citenum{JCP_136_064902} and ~\citenum{JCP_143_024901} provide all the essential technical details necessary in order to derive the expression for the free energy of compressible semiflexible linear telechelic polymer melts considered in the present paper.

The Helmholtz free energy $f$ per lattice site of a semiflexible telechelic polymer melt is conveniently expressed as the sum of the free energy $f_o$ of the hypothetical reference system in the absence of sticky interactions and the free energy contribution $f_s$ arising from the sticky interactions,
\begin{equation}
	f=f_o+f_s.
\end{equation}
By construction, $f_o$ is independent of $\epsilon_s$ and $E_s$, while $f_s$ depends on these energy parameters as well as the other parameters of the model.

The LCT~\cite{ACP_103_335, JCP_141_044909} yields the Helmholtz free energy $f_o$ of a semiflexible linear polymer melt in the following form,
\begin{equation}
	\beta f_o=\beta f_o^{mf}-\sum_{i=1}^6C_i\phi^i,
\end{equation}
where $\beta=1/(k_BT)$ with $k_B$ being Boltzmann's constant and $T$ designating the absolute temperature. The first term $\beta f_o^{mf}$ in Eq. (2) represents the zeroth-order mean-field contribution and appears as
\begin{eqnarray}
	\beta f_o^{mf}=&&\frac{\phi}{M}\ln\left(\frac{2\phi}{zM}\right)+\phi\left(1-\frac{1}{M}\right)\nonumber\\
	&&
	+ (1-\phi)\ln(1-\phi)-\phi \frac{N_{2}}{M}\ln(z_b),
\end{eqnarray}
where $N_2$ is the number of runs of two consecutive bonds in a single chain, and $z_b=(z_p-1)\exp(-\beta E_b)+1$ with $z_p=z/2$. The second term in Eq. (2) is due to corrections to the zeroth-order mean-field free energy $\beta f_o^{mf}$ arising from the short range correlations possible for clusters containing at most four consecutive bonds, and the coefficients $C_i$ $(i=1, ..., 6)$ are presented as a polynomial in power of $\phi$ and generally depends on $z$, $T$, $\epsilon$, $E_b$, and a set of counting indices $u_i=N_i/M$ $(i=1, ..., 4)$, where the counting factor $N_i$ denotes the number of runs of $i$ consecutive bonds in a single chain and is simply equals to $N_i=M-i$ for linear chains. Reference~\citenum{JCP_143_024901} provides explicit expressions for $C_i$ $(i=1, ..., 6)$ for a melt of semiflexible linear chains.

As shown in Refs.~\citenum{JCP_136_064902} and ~\citenum{JCP_143_024901}, the sticky contribution $f_s$ is derived as the series,
\begin{equation}
	\beta f_s=\beta f_s^{mf}-\sum_{i=1}^4Y_iy^i,
\end{equation}
in the density $y$ of sticky bonds, which is defined as the ratio of the number of sticky bonds in the system to the total number of lattice sites. The leading zeroth-order mean-field contribution from sticky interactions $\beta f_s^{mf}$ to the free energy emerges as
\begin{eqnarray}
	\beta f_s^{mf}=&&-\phi x\ln(\phi x)+(\phi x-2y)\ln(\phi x-2y)\nonumber\\
	&&
	+y\left[1+\ln(2y/z)+\beta\epsilon_s\right],
\end{eqnarray}
where $x=2/M$ denotes the fraction of stickers in a single chain. Similarly, the second term in Eq. (4) is due to corrections to the zeroth-order mean-field contribution $\beta f_s^{mf}$ arising from short range correlations in clusters of at most four consecutive bonds containing at least one sticky bond. Appendix A provides explicit expressions for $Y_i$ $(i=1,...,4)$. Notice that the coefficients $Y_i$ $(i=1,...,4)$ now depend on $E_s$ because of the stiffness introduced by the sticky bonds. When $E_s$ vanishes, the theory reduces identically to that presented in Ref.~\citenum{JCP_143_024901}. 

The LCT~\cite{JCP_136_064902} employs the maximum term method to determine the variable $y$ in Eqs. (4) and (5),
\begin{eqnarray}
	\left. \frac{\partial (\beta f_s)}{\partial y}\right|_{T, \phi}=0.
\end{eqnarray}
The solution $y^{\ast}$ of Eq. (6) denotes the equilibrium concentration of the sticky bonds under given thermodynamic conditions. Substituting $y^{\ast}$ into Eqs. (1),  (4), and (5) leads to the final expression for the free energy $f$ of a semiflexible linear telechelic melt, 
\begin{eqnarray}
	\beta f=&&\beta f_o-\phi x\ln(\phi x)+(\phi x-2y^{\ast})\ln(\phi x-2y^{\ast})\nonumber\\
	&&
	+y^{\ast}\left[1+\ln(2y^{\ast}/z)+\beta\epsilon_s\right]-\sum_{i=1}^4Y_i(y^{\ast})^i.
\end{eqnarray}

Evidently, the quantity $y^{\ast}$ depends on all molecular and thermodynamic parameters (such as $T$, $\phi$, $M$, $\epsilon$, $E_b$, $E_s$, and $\epsilon_s$) and plays a central role in the LCT in determining the thermodynamic properties of telechelic polymers. Because the present model assumes that each sticker is mono-functional, the filling fraction of the stickers participating in sticky interactions is simply $2y^{\ast}$ for any given thermodynamic conditions subject to the upper limit for $y^{\ast}$ is $y_{max}^{\ast}=\phi/M$. While the current version of the LCT provides no explicit information regarding the concentration of sticky bonds in the cyclic clusters, cyclic clusters may form.

\section{Results and discussion}

This section presents  illustrative calculations describing the thermodynamics of self-assembly and glass-formation in the model of semiflexible telechelic polymer melts. Special focus is placed on examining the influence of the stiffness of sticky bonds on the average degree and transition temperature of self-assembly, followed by a discussion of the influence on glass-formation of both the sticky and bending rigidity parameters. As in previous work,~\cite{JCP_136_064903, JCP_136_194902, JCP_143_024902} all computations in the present paper are obtained by taking the lattice coordination number as $z=6$. 

\subsection{Influence of stiffness of the sticky bonds on the average degree of self-assembly}

The analysis begins by exhibiting the substantial impact that the stiffness of the sticky bonds may exert upon the average degree of self-assembly in telechelic polymer melts. As derived in Ref.~\citenum{JCP_136_194902}, the average degree $<N>$ of self-assembly for the present lattice model is given by
\begin{equation}
	<N>\approx \frac{1}{1-\Phi},
\end{equation}
where $\Phi=y^*/y^*_{max}$ is the order parameter of self-assembly. The concentration $y^{\ast}$ of the sticky bonds is thus directly related to the average degree of self-assembly. Hence, we now focus on the dependence of $y^{\ast}$ on the sticky bending energy. 

\begin{figure}[tb]
	\centering
	\includegraphics[angle=0,width=0.45\textwidth]{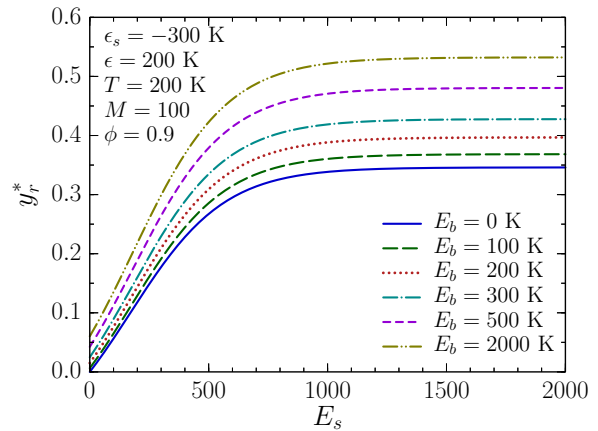}
	\caption{Dependence of the ratio $y_r^{\ast}=(y^{\ast}-y_0^{\ast})/y_0^{\ast}$ on the sticky bending rigidity parameter $E_s$ for various ordinary bending rigidity parameters $E_b$. The computations are performed for a melt of linear telechelic chains, where the polymer filling fraction is $\phi=0.9$, the molecular weight of an individual unassociated chain is $M=100$, the cohesive interaction energy parameter is $\epsilon=200$ K, and the sticky interaction energy parameter is $\epsilon_s=-300$ K. The temperature is fixed to be $T=200$ K.}
\end{figure}

Our previous work~\cite{JCP_143_024902} indicates that the quantitative effect of the ordinary bending rigidity parameter $E_b$ on $y^{\ast}$ is quite small for a wide range of polymer filling fractions and temperatures when the sticky bonds are fully flexible (i.e., $E_s=0$ K). Therefore, the ratio $y_r^{\ast}$ is introduced to measure the relative change of $y^{\ast}$ with increasing the bending rigidity parameters and defined as
\begin{equation}
	y_r^{\ast}=\frac{y^{\ast}-y_0^{\ast}}{y_0^{\ast}},
\end{equation}
where $y_0^{\ast}$ is the value of $y^{\ast}$ for fully flexible chains (i.e., $E_b=E_s=0$ K). One advantage of using such a ratio is that the sign of $y_r^{\ast}$ directly indicates whether chain stiffness promotes ($y_r^{\ast}>0$) or opposes ($y_r^{\ast}<0$) self-assembly. Figure 3 displays $y_r^{\ast}$ as a function of the sticky bending rigidity parameter $E_s$ for various ordinary bending rigidity parameters $E_b$, when all other parameters of the model remain constant. Using this parameter set, Fig. 3 displays $y_r^{\ast}$ as first increasing with $E_s$ for each $E_b$ and then reaching a constant for sufficiently large $E_s$. Figure 3 further reveals that the quantitative influence of $E_s$ on $y^{\ast}$ is much stronger than that of $E_b$. For instance, $y^{\ast}$ increases by nearly $35\%$ for the parameter set used in Fig. 3 when $E_s$ is elevated from $0$ K to $2000$ K at $E_b=0$ K, while increasing $E_b$ from $0$ K to $2000$ K at $E_s=0$ K leads to a much smaller (about $6\%$) increase in $y^{\ast}$. This analysis thus implies that the average degree of self-assembly is strongly influenced by the stiffness of the sticky bonds in telechelic polymers.

\begin{figure}[tb]
	\centering
	\includegraphics[angle=0,width=0.45\textwidth]{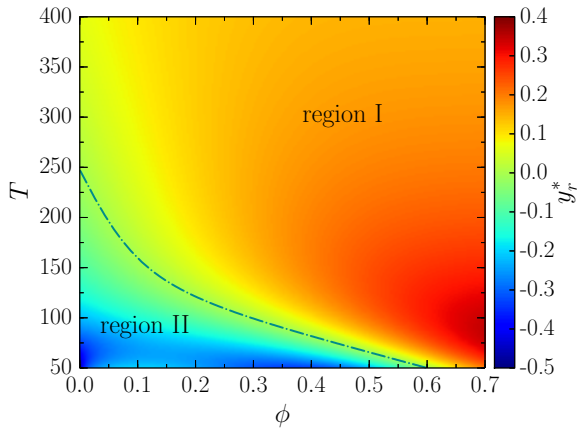}
	\caption{Contour plot of $y_r^{\ast}$ in the $\phi$-$T$ plane. The dashed-dotted line denotes the boundary demarking states with $y^{\ast}=0$. Chain stiffness promotes or opposes self-assembly in region I or II, respectively. The computations are performed for a melt of linear telechelic chains, where the molecular weight is $M=100$, the cohesive interaction energy parameter is $\epsilon=200$ K, the sticky interaction energy parameter is $\epsilon_s=-100$ K, and the ordinary and sticky bending rigidity parameters are $E_b=2000$ K and $E_s=150$ K.}
\end{figure}

One interesting feature exhibited by the lattice model of telechelic polymers is that chain stiffness can either promote or oppose self-assembly, depending on the thermodynamic conditions considered. Specifically, our previous work~\cite{JCP_143_024902} reveals that the average degree of self-assembly in the model of telechelic polymer melts with fully flexible sticky bonds diminishes with increasing the ordinary bending rigidity parameter $E_b$ when both $\phi$ and $T$ are sufficiently low. This feature is demonstrated here to persist in the model of telechelic polymers with semiflexible sticky bonds. As an illustration, Fig. 4 displays the contour plot of $y_r^{\ast}$ in the $\phi$-$T$ plane, where the ordinary and sticky bending rigidity parameters are $E_b=2000$ K and $E_s=150$ K, respectively. As can be seen, chain stiffness promotes self-assembly for systems represented in the $\phi$-$T$ plane where either $\phi$ or $T$ is high (termed region I), while self-assembly can be suppressed by chain stiffness when both $\phi$ and $T$ are sufficiently low (termed region II).

Our previous work~\cite{JCP_143_024902} invokes a Flory-Huggins (FH) type theory~\cite{JCP_136_244904} for the competition between the formation of rings versus linear clusters in order to provide a possible rationale for the opposite variations with chain stiffness of self-assembly in different regions. The present LCT provides no information concerning the formation of rings, as noted in Sec. II. In particular, the FH type theory~\cite{JCP_136_244904} predicts that linear clusters form more easily than rings at high $\phi$. At low $\phi$, however, rings predominate over linear clusters at low $T$, whereas the opposite situation ensues at high $T$, a behavior that arises because the extra bond energy gained upon ring closure outweighs the entropy loss upon ring closure as $T$ decreases. Therefore, the formation of rings is expected to be favored when both $\phi$ and $T$ are low. The trend of forming linear clusters is thus enhanced by chain stiffness because of a diminished probability of ring closure as the chains stiffen. Moreover, the formation of sticky bonds between different chains becomes less sterically hindered as the chains are stiffer. Therefore, chain stiffness promotes self-assembly under the conditions where linear clusters predominate. Meanwhile, we conjecture that if cyclic clusters predominate, the gain in sticky bonds due to the enhancement of linear clusters induced by chain stiffness cannot compensate for the loss of sticky bonds generated by the reduction in ring formation due to the stiffness,  and consequently, chain stiffness opposes the self-assembly. The above explanation, if confirmed, e.g., by computer simulations, likewise applies for the present model with semiflexible sticky bonds.

\begin{figure}[tb]
	\centering
	\includegraphics[angle=0,width=0.45\textwidth]{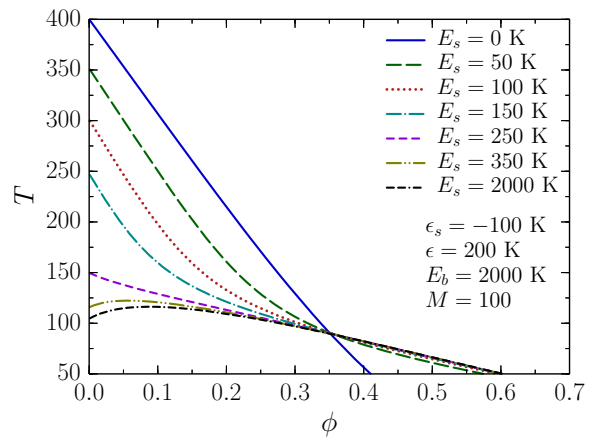}
	\caption{Dependence of the boundary line in the $\phi$-$T$ plane on the sticky bending rigidity parameter $E_s$. The computations are performed for a melt of linear telechelic chains, where the molecular weight is $M=100$, the cohesive interaction energy parameter is $\epsilon=200$ K, the sticky interaction energy parameter is $\epsilon_s=-100$ K, and the ordinary bending rigidity parameter is $E_b=2000$ K.}
\end{figure}

Following our previous analysis,~\cite{JCP_143_024902} a boundary (shown as a dashed-dotted line in Fig. 4) in the $\phi$-$T$ plane with $y_r^{\ast}=0$ separates two regions with opposite dependences of $y^{\ast}$ on chain stiffness. Our previous work~\cite{JCP_143_024902} examines the variation of the boundary with various molecular parameters (such as $\epsilon_s$, $E_b$, $M$, and $\epsilon$), indicating that the boundary is insensitive to $\epsilon_s$ but depends on other parameters. For instance, the area of region II shrinks slightly with increasing $E_b$ or $M$ and saturates for sufficiently large $E_b$ or $M$, while elevating $\epsilon$ leads to a dramatic increase in the area of region II in the $\phi$-$T$ plane. The above trends are found to apply for the present model with $E_s>0$ K (data not shown). The influence of the sticky bending rigidity parameter $E_s$ on the boundary in the $\phi$-$T$ plane is presented in Fig. 5, which indicates that the boundary strongly depends on $E_s$. In particular, the temperatures marking the boundary significantly descend with increasing $E_s$ in the low $\phi$ regime but ascend in the high $\phi$ regime. Interestingly, these boundary lines intersect at a common point with $\phi=0.35$ and $T=90.7$ K for various $E_s$. Unfortunately, the physical significance for the presence of such a point is unclear at present. 

\subsection{Influence of stiffness of the sticky bonds on the self-assembly transition}

The transition temperature $T_p$ of self-assembly is an important quantity in the thermodynamic description of self-assembly. As implied by the FH type theories of self-assembly,~\cite{JCP_119_12645} one common definition for $T_p$ employs the temperature variation of the order parameter $\Phi$ of self-assembly. Specifically, $T_p$ is identified with the temperature at which $\Phi(T, \phi=\text{const})$ exhibits an inflection point as a function of $T$, i.e., the temperature at which the second derivative of $\Phi$ with respect to $T$ vanishes,
\begin{equation}
	\left.\frac{\partial^2 \Phi}{\partial T^2}\right|_{\phi}=\left.\frac{\partial^2 y^*}{\partial T^2}\right|_{\phi}=0.
\end{equation}

\begin{figure}[tb]
	\centering
	\includegraphics[angle=0,width=0.45\textwidth]{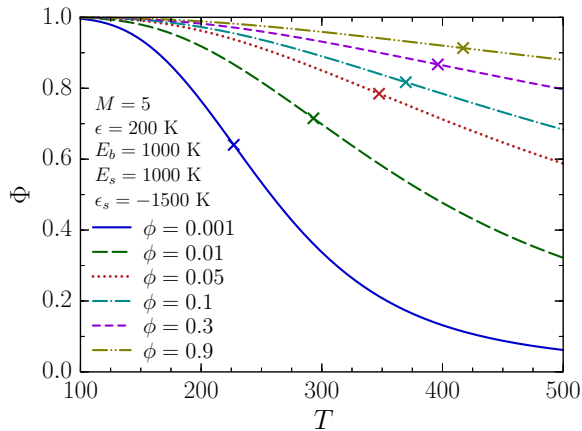}
	\caption{Temperature variation of the order parameter $\Phi=y^*/y^*_{max}$ of self-assembly for various polymer filling fractions $\phi$. Crosses indicate the positions of the inflection points of the curves. The computations are performed for a melt of linear telechelic chains, where the molecular weight is $M=5$, the cohesive interaction energy parameter is $\epsilon=200$ K, the sticky interaction energy parameter is $\epsilon_s=-1500$ K, and the ordinary and sticky bending rigidity parameters are $E_b=1000$ K and $E_s=1000$ K.}
\end{figure}

\begin{figure}[tb]
	\centering
	\includegraphics[angle=0,width=0.45\textwidth]{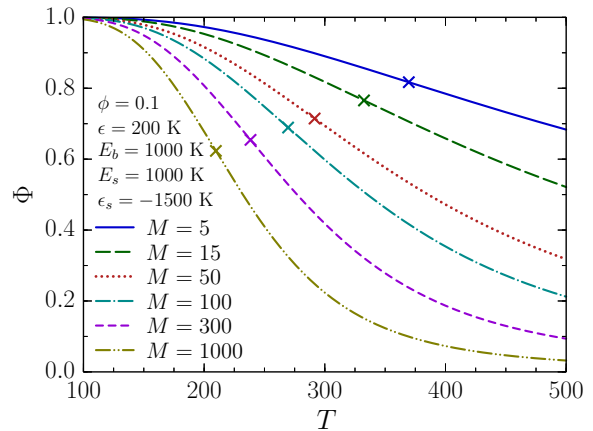}
	\caption{Temperature variation of the order parameter $\Phi=y^*/y^*_{max}$ of self-assembly for various molecular weights $M$. Crosses indicate the positions of the inflection points of the curves. The computations are performed for a melt of linear telechelic chains, where the polymer filling fraction is $\phi=0.1$, the cohesive interaction energy parameter is $\epsilon=200$ K, the sticky interaction energy parameter is $\epsilon_s=-1500$ K, and the ordinary and sticky bending rigidity parameters are $E_b=1000$ K and $E_s=1000$ K.}
\end{figure}

\begin{figure}[tb]
	\centering
	\includegraphics[angle=0,width=0.45\textwidth]{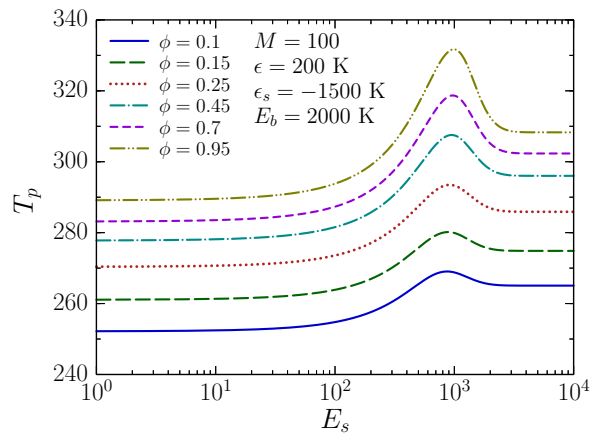}
	\caption{Transition temperature $T_p$ for self-assembly as a function of the sticky bending rigidity parameter $E_s$ for various polymer filling fractions $\phi$. The computations are performed for a melt of linear telechelic chains, where the molecular weight is $M=100$, the cohesive interaction energy parameter is $\epsilon=200$ K, the sticky interaction energy parameter is $\epsilon_s=-1500$ K, and the ordinary bending rigidity parameter is $E_b=2000$ K.}
\end{figure}

\begin{figure}[tb]
	\centering
	\includegraphics[angle=0,width=0.45\textwidth]{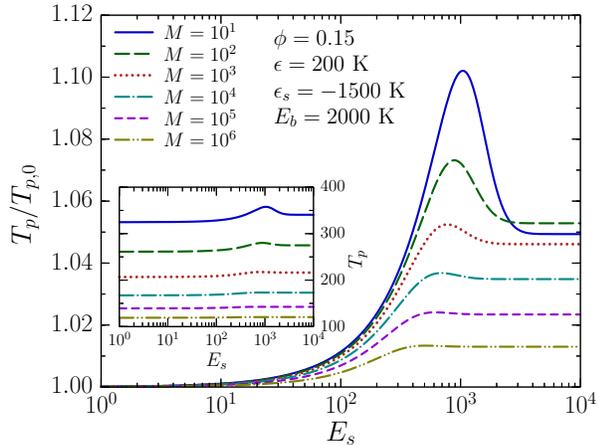}
	\caption{$T_p/T_{p,0}$ as a function of the sticky bending rigidity parameter $E_s$ for various molecular weights $M$, where $T_{p,0}$ designates the value of $T_p$ at $E_s=0$ K for each $M$. The inset depicts $T_p$ as a function of $E_s$ for various $M$. The computations are performed for a melt of linear telechelic chains, where the polymer filling fraction is $\phi=0.15$, the cohesive interaction energy parameter is $\epsilon=200$ K, the sticky interaction energy parameter is $\epsilon_s=-1500$ K, and the ordinary bending rigidity parameter is $E_b=2000$ K.}
\end{figure}

Figures 6 and 7 present, respectively, the temperature variation of $\Phi$ for various $\phi$ and $M$, when the other parameters are fixed. Since the present model considers examples where self-assembly of the telechelic chains is promoted upon cooling, self-assembly flourishes at low $T$ where $\Phi$ may approach unity. Analysis of Figs. 6 and 7 reveals the presence of an inflection point in each curve. Hence, $T_p$ may likewise be identified in the LCT for self-assembling telechelic polymers from the inflection points in $\Phi(T, \phi=\text{const})$, in agreement with previous calculations for fully flexible telechelic polymers.~\cite{JCP_136_194902} Moreover, the self-assembly transition in models of telechelic polymers is found to be very broad, and the broadness of the transition grows with increasing polymer filling fraction $\phi$ or decreasing molecular weight $M$. These trends also accord with those for fully flexible telechelic polymers.~\cite{JCP_136_194902}

Previous work~\cite{JCP_136_194902} extensively examines the dependence on thermodynamic and molecular parameters of the transition temperature $T_p$ for self-assembling telechelic polymers composed of fully flexible chains. For instance, $T_p$ is found to increase with elevating $\phi$ or $|\epsilon_s|$ but decrease with growing $M$. These general trends also remain in the present model of semiflexible telechelic polymers (data not shown). Figure 8 further reveals the strong influence of the sticky bending rigidity parameter $E_s$ on $T_p$. When $\phi$ is held constant, $T_p$ is shown to first grow with $E_s$, display a maximum, then to reduce with $E_s$, and eventually reach a constant for sufficiently large $E_s$. Our calculations also indicate a non-monotonic change of $T_p$ with $E_b$ for fixed $E_s$ (data not shown). Figure 9 further examines how the chain length alters the dependence of $T_p$ on $E_s$. In particular, Fig. 9 presents both $T_p$ and $T_p/T_{p,0}$ as a function of $E_s$ for various $M$, where $T_{p,0}$ is the value of $T_p$ at $E_s=0$ K for each $M$. Evidently, the influence of $E_s$ on $T_p$ progressively weakens with increasing $M$, as expected. Consequently, the non-monotonic variation of $T_p$ with $E_s$ is less evident for larger $M$ and becomes barely detectable for sufficiently large $M$, where $T_p$ indeed depends very weakly on $E_s$; e.g., $T_p$ increases by less than $2\%$ for $M=10^6$ when $E_s$ is elevated from $0$ K to $10^4$ K. Notice that the results in Figs. 8 and 9 are presented for a quite broad range of $E_s$ in order to reach saturation, which clearly only emerges for $E_s>\sim4000$ K. The same consideration applies in the following computations.

Notably, the transition temperatures for semiflexible chains become elevated as compared to those for fully flexible chains, in agreement with our earlier analysis in Fig. 5 that chain stiffness promotes self-assembly for the parameter set considered in Figs. 8 and 9. Therefore, the above results clearly demonstrate the important role of chain stiffness, in particular, the stiffness of the sticky bonds, in the thermodynamic description of telechelic polymers.

\begin{figure}[tb]
	\centering
	\includegraphics[angle=0,width=0.45\textwidth]{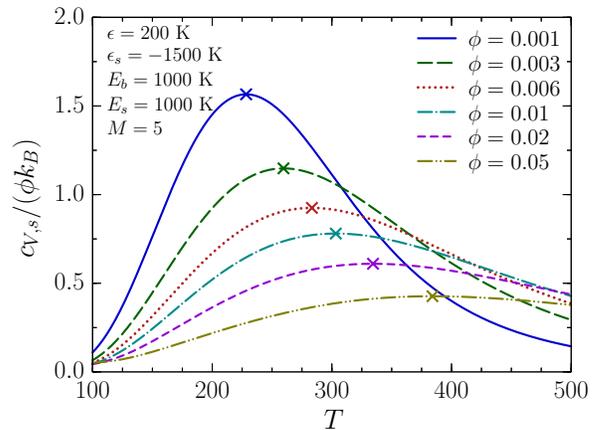}
	\caption{Temperature variation of $c_{V,s} k_B$ normalized by $\phi$ for various polymer filling fractions $\phi$. Crosses indicate the positions of the maxima of the curves. The computations are performed for a melt of linear telechelic chains, where the molecular weight is $M=5$, the cohesive interaction energy parameter is $\epsilon=200$ K, the sticky interaction energy parameter is $\epsilon_s=-1500$ K, and the ordinary and sticky bending rigidity parameters are $E_b=1000$ K and $E_s=1000$ K.}
\end{figure}

\begin{figure}[tb]
	\centering
	\includegraphics[angle=0,width=0.45\textwidth]{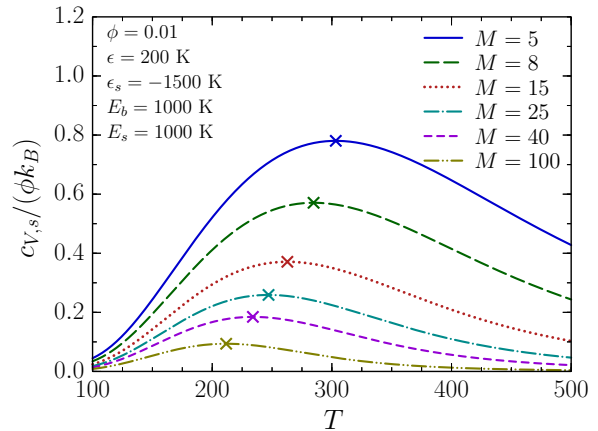}
	\caption{Temperature variation of $c_{V,s} k_B$ normalized by $\phi$ for various molecular weights $M$. Crosses indicate the positions of the maxima of the curves. The computations are performed for a melt of linear telechelic chains, where the polymer filling fraction is $\phi=0.01$, the cohesive interaction energy parameter is $\epsilon=200$ K, the sticky interaction energy parameter is $\epsilon_s=-1500$ K, and the ordinary and sticky bending rigidity parameters are $E_b=1000$ K and $E_s=1000$ K.}
\end{figure}

An alternative identification~\cite{JCP_136_194902} for $T_p$ involves the maximum in the specific heat $c_V(T)$.~\footnote{The use of a lattice model implies that the computed specific heat is devoid of the substantial contributions from molecular vibrations, and therefore, data for $c_V$ should be interpreted with caution.} The specific heat $c_V$ is determined as usual from the second derivative of the Helmholtz free energy $f$ with respect to the inverse temperature $\beta=1/(k_BT)$,
\begin{equation}
	\frac{c_V}{k_B}=-\beta^2\left.\frac{\partial^2 (\beta f)}{\partial \beta^2}\right|_{\phi}.
\end{equation}
Similar to the free energy $f$, $c_V$ for the model of self-assembling telechelic polymers is composed of two separate contributions $c_{V,o}$ and $c_{V,s}$, which arise, respectively, from the reference system and the sticky interactions. These two terms thus appear as
\begin{equation}
	\frac{c_{V,o}}{k_B}=-\beta^2\left.\frac{\partial^2 (\beta f_o)}{\partial \beta^2}\right|_{\phi},
\end{equation}
and
\begin{equation}
	\frac{c_{V,s}}{k_B}=-\beta^2\left.\frac{\partial^2 (\beta f_s)}{\partial \beta^2}\right|_{\phi}.
\end{equation}

Previous work~\cite{JCP_136_194902} demonstrates the presence of a maximum in the temperature dependence of $c_{V,s}$, which, in turn, provides a definition for the transition temperature $T_p$ of self-assembly in telechelic polymers composed of fully flexible chains. This identification is now tested for the model of semiflexible telechelic polymers. We focus on the regime of low $\phi$ since the presence of peaks in $c_{V,s}(T)$ seems to be more pronounced at lower $\phi$. Figures 10 and 11 display the temperature variation of $c_{V,s}/k_B$ (normalized by $\phi$) for various $\phi$ and $M$, respectively. A maximum in $c_{V,s}(T)$ appears in each curve, thereby allowing for an alternative determination of $T_p$ from $c_{V,s}(T)$. In particular, the self-assembly transition broadens and the maximum of $c_{V,s}$ shifts to high temperatures as $\phi$ increases or $M$ decreases, supporting our earlier results for the self-assembly transition from the inflection points in $\Phi(T, \phi=\text{const})$. While the transition temperatures from both methods quantitatively differ, the general trends of $T_p$ as determined either from $\Phi$ or $c_{V,s}$ track each other when individual molecular parameters are varied.

\subsection{Glass-formation in the lattice model of linear telechelic polymer melts}

One substantial benefit of the LCT for describing the thermodynamic properties of semiflexible telechelic polymers lies in the fact that glass-formation in such systems can be addressed by combining the LCT with the AG relation,~\cite{JCP_43_139} thereby extending the GET~\cite{ACP_137_125} to the self-assembling telechelic polymers. Freed~\cite{JCP_141_141102} generalizes transition state theory to account for collective barrier-crossing events, thereby providing a firm theoretical foundation for the principal assumptions of the AG theory, so the AG model is taken as well established for polymer melts. This generalization enables investigations of the phenomenon of glassy behavior that is influenced by self-assembly. This section provides basic information concerning the GET, followed by a brief discussion of glass-formation in the model of telechelic polymers. In particular, the stiffness of the sticky bonds is demonstrated to significantly influence glass-formation in telechelic polymers.

\begin{figure}[tb]
	\centering
	\includegraphics[angle=0,width=0.45\textwidth]{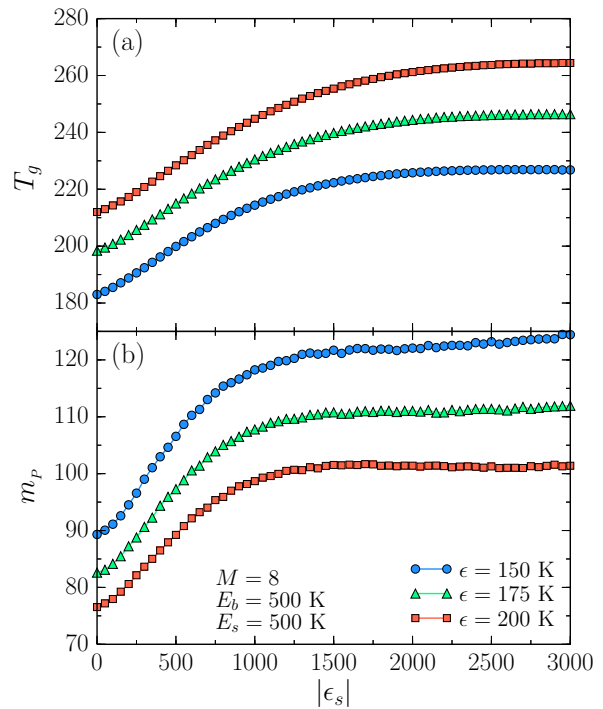}
	\caption{(a) Glass transition temperature $T_g$ and (b) isobaric fragility parameter $m_{_P}$ as a function of the absolute sticky interaction energy parameter $|\epsilon_s|$ for various cohesive interaction energy parameters $\epsilon$. The computations are performed for a melt of linear telechelic chains at a constant pressure of $P=0.101~325$ MPa (i.e., $1$ atm), where the molecular weight is $M=8$, and the ordinary and sticky bending rigidity parameters are $E_b=500$ K and $E_s=500$ K.}
\end{figure}

\begin{figure}[tb]
	\centering
	\includegraphics[angle=0,width=0.45\textwidth]{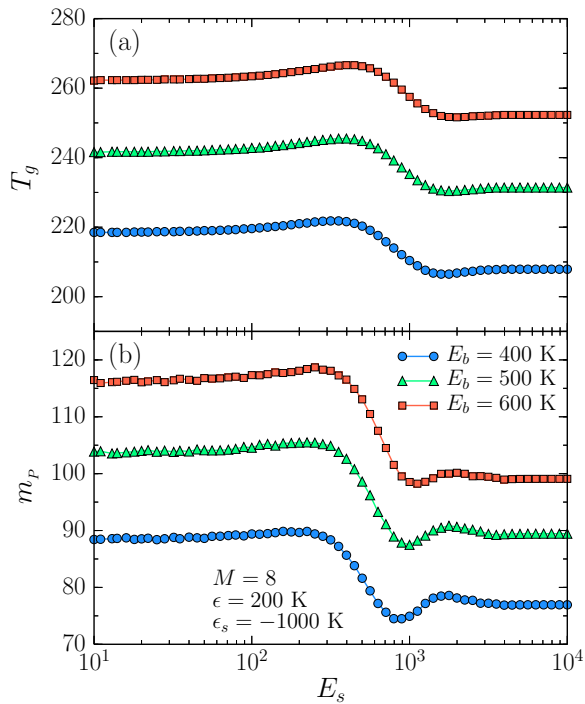}
	\caption{(a) Glass transition temperature $T_g$ and (b) isobaric fragility parameter $m_{_P}$ as a function of the sticky bending rigidity parameter $E_s$ for various ordinary bending rigidity parameters $E_b$. The computations are performed for a melt of linear telechelic chains at a constant pressure of $P=0.101~325$ MPa (i.e., $1$ atm), where the molecular weight is $M=8$, the cohesive interaction energy parameter is $\epsilon=200$ K, and the sticky interaction energy parameter is $\epsilon_s=-1000$ K.}
\end{figure}

\begin{figure}[tb]
	\centering
	\includegraphics[angle=0,width=0.45\textwidth]{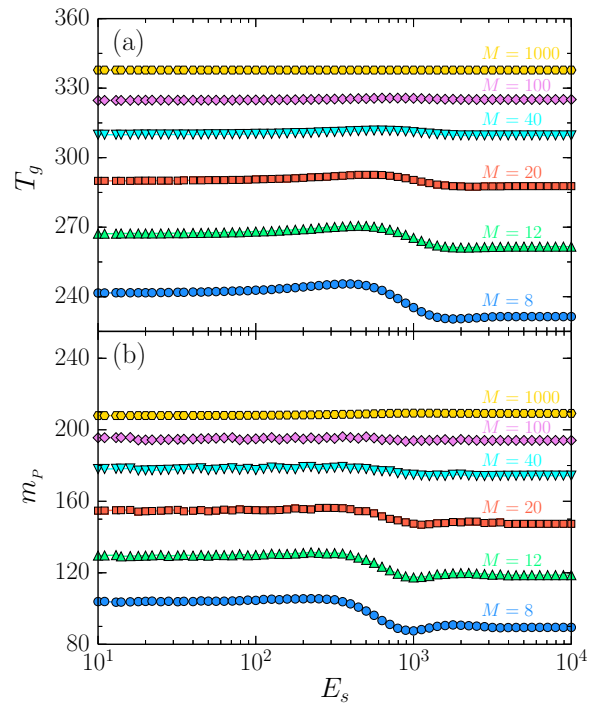}
	\caption{(a) Glass transition temperature $T_g$ and (b) isobaric fragility parameter $m_{_P}$ as a function of the sticky bending rigidity parameter $E_s$ for various molecular weights $M$. The computations are performed for a melt of linear telechelic chains at a constant pressure of $P=0.101~325$ MPa (i.e., $1$ atm), where the cohesive interaction energy parameter is $\epsilon=200$ K, the sticky interaction energy parameter is $\epsilon_s=-1000$ K, and the ordinary bending rigidity parameter is $E_b=500$ K.}
\end{figure}

Polymer glass-formation is treated in the GET as a broad transition with four characteristic temperatures.~\cite{ACP_137_125} These characteristic temperatures are obtained first by evaluating the configurational entropy density (defined by $s_c=-\partial f/\partial T |_{\phi}$, i.e., the configurational entropy per lattice site~\cite{JCP_119_5730, JCP_141_234903, paper_note1}) at constant pressure ($P$). The temperature variation of the configurational entropy density $s_c(T)$ exhibits features that enable the direct determination of three characteristic temperatures of glass formation, namely, the onset temperature $T_A$ which signals the onset of non-Arrhenius behavior of the structural relaxation time and which is found from the maximum in $s_c(T)$, the ideal glass transition temperature $T_o$ where $s_c$ extrapolates to zero, and the crossover temperature $T_c$ which separates two temperature regimes with qualitatively different dependences of the structural relaxation time on temperature and which is evaluated from the inflection point in $Ts_c(T)$. The glass transition temperature $T_g$ is determined by calculating the structural relaxation time $\tau_{\alpha}$ via the AG relation,~\cite{JCP_43_139}
\begin{equation}
	\tau_{\alpha}=\tau_\infty\exp[\beta\Delta\mu s_c^\ast/s_c(T)],
\end{equation}
where $\tau_\infty$ is the high temperature limit of the relaxation time, $\Delta\mu$ is the high temperature activation free energy, and $s_c^\ast$ is the high temperature limit of $s_c(T)$ [identified by $s_c^\ast=s_c(T_A)$  in the GET]. $\tau_\infty$ is set to be $10^{-13}$ s in the GET as a typical value for polymers.~\cite{PRE_67_031507} Motivated by experimental data for the crossover temperature of various glass-formers,~\cite{PRE_67_031507} the GET estimates the high temperature activation energy from the empirical relation $\Delta\mu=6k_BT_c$.~\cite{ACP_137_125} The GET then identifies $T_g$ as the temperature at which $\tau_{\alpha}=100$ s. Likewise, the isobaric fragility parameter $m_{_P}$ is determined from the standard definition,~\cite{Science_267_1924}
\begin{equation}
	m_{_P}=\left.\frac{\partial \log (\tau_{\alpha})}{\partial (T_g/T)}\right|_{P, T=T_g}.
\end{equation}
Illustrative computations of characteristic temperatures and fragility parameters and more details concerning the GET can be found in previous works (e.g., see Refs.~\citenum{ACP_137_125} and \citenum{Mac_47_6990}).

The calculations are performed at a constant pressure of $P=0.101~325$ MPa (i.e., $1$ atm) and use the common parameters $z=6$ and $V_{\text{cell}}=(2.7)^3$\AA{}$^3$. Here, $V_{\text{cell}}$ is introduced to describe the volume of a single lattice site, a parameter that is required in order to express the pressure in real units. A low molecular weight of $M=8$ is first chosen for our illustrative calculations because the quantitative influence of the sticky interaction energy on glass-formation becomes more significant for smaller $M$. The influence of $M$ on glass-formation in the model of telechelic melts is then briefly discussed.

Figure 12 displays the dependence of $T_g$ and $m_{_P}$ on the absolute sticky interaction energy parameter $|\epsilon_s|$ for various microscopic cohesive interaction energy parameters $\epsilon$. Both $T_g$ and $m_{_P}$ increase with the sticky interaction strength and tend to saturate for sufficiently strong sticky interactions. These trends are quite understandable since an increase in $|\epsilon_s|$ elevates the ``effective'' molecular weight due to increases in the average degree of self-assembly. Growing molecular weight usually leads to increases in both $T_g$ and $m_{_P}$ in polymer melts (see Fig. 14), and hence, both $T_g$ and $m_{_P}$ are expected to become larger for longer averaged chain lengths induced by stronger sticky interactions. In addition, $T_g$ appears to saturate at higher $|\epsilon_s|$ than $m_{_P}$. Apparently, when $|\epsilon_s|$ becomes large for a telechelic melt, $T_g$ or $m_{_P}$ can be identical to that for a polymer melt with $\epsilon_s=0$ K but with a higher $M$, defining an effective molecular weight for telechelics with a given sticky energy. This complicated point, however, is not relevant to the present paper. Figure 12 also indicates that a larger $\epsilon$ results in a higher $T_g$ but a lower $m_{_P}$ for a fixed $|\epsilon_s|$, trends that are the same as those in polymer melts lacking sticky interactions.~\cite{JCP_141_234903, Mac_47_6990, Mac_48_2333, JCP_131_114905}

The influence of bending rigidity parameters on glass-formation is examined in Fig. 13, which exhibits the dependence of both $T_g$ and $m_{_P}$ on the sticky bending rigidity parameter $E_s$ for various ordinary bending rigidity parameters $E_b$. As for polymer melts without sticky interactions,~\cite{JCP_141_234903, Mac_47_6990, Mac_48_2333, JCP_131_114905} elevating $E_b$ causes both $T_g$ and $m_{_P}$ to grow in telechelic polymers.~\footnote{While Fig. 13 displays the calculations for a relatively narrow range of $E_b$ from $400$ to $600$ K, the trends of elevating $T_g$ and $m_{_P}$ with $E_b$ holds for a wide range of $E_b$, where the chains vary from very flexible to very stiff. Of course, both $T_g$ and $m_{_P}$ will saturate for sufficiently large $E_b$ when reaching the stiff chain limit.} Turning to the role of the sticky bending energy, Fig. 13 reveal somewhat complicated variations of $T_g$ and $m_{_P}$ with $E_s$. For instance, both $T_g$ and $m_{_P}$ first grow slightly upon increasing $E_s$ for fixed $E_b$, drop for intermediate values of $E_s$, then become increased again, and eventually plateau for sufficiently large $E_s$. Figure 14 further explors the dependence of $T_g$ and $m_{_P}$ on $E_s$ for various $M$. Apparently, a larger $M$ leads to a weaker dependence of $T_g$ and $m_{_P}$ on $E_s$ and the influence of $E_s$ on polymer glass-formation is almost negligible for sufficiently large $M$, results that are in accord with expectations since the sticky contributions to the free energy decrease considerably with growing $M$ in the present model. Hence, the non-monotonic dependence of $T_g$ and $m_{_P}$ on $E_s$ is less evident for larger $M$ and becomes nearly invisible for sufficiently large $M$. While the non-monotonic behavior shown in Figs. 13 and 14 remains to be fully understood, perhaps requiring simulations, our calculations clearly demonstrate that the stiffness of the sticky bonds greatly affects glass-formation in the model of telechelic polymers, at least for short chains.

\section{Summary}

Despite the fact that telechelic polymers can be used as building blocks for designing important materials, a predictive molecular theory has been slow to develop for describing the influence of key molecular factors on the physical properties of such systems. Currently available theories~\cite{Mac_28_1066, Mac_28_7879, JCP_110_1781, Mac_33_1425, Mac_33_1443, JCP_119_6916, Lan_20_7860, JPSB_45_3285, JCP_131_144906, JPCB_114_12298} and simulations~\cite{Mac_20_1999, JCP_110_6039, EPL_59_384, Polymer_45_3961, JPSB_43_796, PRL_96_187802, PRL_109_238301, JCP_126_044907, JPCM_20_335103, Mac_47_4118, Mac_47_6946, JCP_143_243117, SM_2016} for telechelic polymers traditionally use highly coarse-grained models that represent the assembling molecular species as a structureless entity. The LCT~\cite{JCP_87_7272, Mac_24_5076, ACP_103_335, APS_183_63} for the thermodynamics of polymer systems instead employs an intermediate level of coarse-grained models that retain essential features of molecular structure and interactions in polymer fluids. Hence, an extension of the LCT to associating telechelic polymers provides a promising theoretical tool for establishing the relation between the molecular structure dependent interaction parameters and the thermodynamic properties of telechelic polymers.

The original LCT for telechelic polymers focuses on the model of fully flexible linear chains,~\cite{JCP_136_064902} in part, because of the great algebraic complexity. Our recent extension of theory~\cite{JCP_143_024901} includes a description of chain semiflexibility by introducing a bending energy penalty whenever a pair of consecutive bonds from the same chains lies along orthogonal directions, but the sticky bonds are treated as being fully flexible. However, the sticky bonds must possess a degree of stiffness due to steric interations that limit relative distances and/or angles of the sticky bonds in real telechelic polymers, thus prompting the present investigation of the influence of stiffness of the sticky bonds on the self-assembly and thermodynamics of telechelic polymers by employing a further extension of the LCT.

Our illustrative calculations indicate that the stiffness of the sticky bonds significantly influences the self-assembly and thermodynamics of telechelic polymers. In particular, previous work~\cite{JCP_143_024902} for telechelic polymers with fully flexible sticky bonds indicates that the average degree of self-assembly is elevated by chain stiffness when either the polymer filling fraction or the temperature is high, but becomes reduced as the chains stiffen when both the polymer filling fraction and temperature are low, and these general trends likewise occur in telechelic polymers with semiflexible sticky bonds. The transition temperature for self-assembly depends non-monotonically on the stiffness of the sticky bonds. Moreover, the present extension of the LCT enables the investigation of glass-formation in telechelic polymers by generalizing the GET to telechelic polymers. The sticky interactions and the stiffness of the sticky bonds emerge from the theory as important molecular factors for tailoring the properties of glass-formation in systems of associating telechelic polymers, at least for short chains.

While the present work only considers linear chains with two mono-functional groups at the chain ends, this theoretical development represents an intermediate step with the development of important extensions of the LCT to describe chains with monomer units possessing specific structures and/or multi-functional stickers.

\begin{acknowledgments}
This work is supported by the National Science Foundation (NSF) Grant No. CHE-1363012.
\end{acknowledgments}

\appendix

\section{Summary of the coefficients that appear in contributions to the free energy arising from the sticky interactions}

The coefficients $Y_i$ $(i=1,...,4)$ that appear in $\beta f_s$ are organized in powers of the polymer filling fraction $\phi$,
\begin{equation}
	Y_i=\sum_{j=0}^{j_{max}}Y_{i,j}\phi^j,
\end{equation}
where $j_{max}=5$, $4$, $2$, and $0$ for $i=1$, $2$, $3$, and $4$, respectively. The explicit expressions for $Y_{i,j}$ are

\begin{subequations}
	\begin{eqnarray}
	Y_{1,0}=&&-\frac{2g_s}{z}-\frac{g_s^2}{z^2}+\frac{N_{2e}(2-g_b-3g_s+g_s^2-g_bg_s^2)}{z^2}\nonumber\\
	&&
	+\frac{N_{3e}(4-6g_b-2g_s+2g_b^2+2g_bg_s-2g_b^2g_s)}{z^2}\nonumber\\
	&&
	+\left(1+\frac{2N_{2e}g_bg_s}{z}\right)(\beta\epsilon),
	\end{eqnarray}
	
	\begin{eqnarray}
	Y_{1,1}=&&\frac{2u_1}{z}+\frac{2u_1(g_s^2-2g_s)+4(u_1N_{2e}+u_2)g_bg_s}{z^2}\nonumber\\
	&&
	+\frac{2u_3g_b^2}{z^2}+\left[-2+\frac{4u_1(1-2g_s)-4N_{2e}g_bg_s}{z}\right.\nonumber\\
	&&
	-\left.\frac{4u_2g_b}{z}\right](\beta\epsilon)+(2g_s-1+2u_1)(\beta\epsilon)^2,
	\end{eqnarray}
	
	\begin{eqnarray}
	Y_{1,2}=&&\frac{u_1^2(2-8g_s)-8u_1u_2g_b}{z^2}+\left[1+\frac{12u_1^2}{z}\right.\nonumber\\
	&&
	+\left.\frac{8u_1(2g_s-1)+8u_2g_b+2N_{2e}g_bg_s}{z}\right](\beta\epsilon)\nonumber\\
	&&
	+(7/2-6g_s-12u_1)(\beta\epsilon)^2,
	\end{eqnarray}
	
	\begin{eqnarray}
	Y_{1,3}=&&\frac{8u_1^3}{z^2}+\left[\frac{4u_1(1-2g_s)-4u_2g_b-24u_1^2}{z}\right](\beta\epsilon)\nonumber\\
	&&
	+(6g_s-4+24u_1)(\beta\epsilon)^2,
	\end{eqnarray}
	
	\begin{eqnarray}
	Y_{1,4}=&&\left(\frac{12u_1^2}{z}\right)(\beta\epsilon)+(3/2-2g_s-20u_1)(\beta\epsilon)^2,\nonumber\\
	&&
	\end{eqnarray}
	
	\begin{eqnarray}
	Y_{1,5}=(6u_1)(\beta\epsilon)^2,
	\end{eqnarray}
\end{subequations}

\begin{subequations}
	\begin{eqnarray}
	Y_{2,0}=&&\frac{1}{z}+\frac{4N_{2e}g_bg_s-4g_s+6g_s^2}{z^2}
	+\left(\frac{2-8g_s}{z}\right)(\beta\epsilon)\nonumber\\
	&&
	+(\beta\epsilon)^2,
	\end{eqnarray}
	
	\begin{eqnarray}
	Y_{2,1}=&&\frac{u_1(2-16g_s)-4u_2g_b}{z^2}
	+\left(\frac{16g_s-4+12u_1}{z}\right)(\beta\epsilon)\nonumber\\
	&&
	-6(\beta\epsilon)^2,
	\end{eqnarray}
	
	\begin{eqnarray}
	Y_{2,2}=&&\frac{12u_1^2}{z^2}+\left(\frac{2-8g_s-24u_1}{z}\right)(\beta\epsilon)
	+12(\beta\epsilon)^2,\nonumber\\
	&&
	\end{eqnarray}
	
	\begin{eqnarray}
	Y_{2,3}=\left(\frac{12u_1}{z}\right)(\beta\epsilon)-10(\beta\epsilon)^2,
	\end{eqnarray}
	
	\begin{eqnarray}
	Y_{2,4}=3(\beta\epsilon)^2,
	\end{eqnarray}
\end{subequations}

\begin{subequations}
	\begin{eqnarray}
	Y_{3,0}=\frac{2/3-8g_s}{z^2}+\left(\frac{4}{z}\right)(\beta\epsilon),
	\end{eqnarray}
	
	\begin{eqnarray}
	Y_{3,1}=\frac{8u_1}{z^2}-\left(\frac{8}{z}\right)(\beta\epsilon),
	\end{eqnarray}
	
	\begin{eqnarray}
	Y_{3,2}=\left(\frac{4}{z}\right)(\beta\epsilon),
	\end{eqnarray}
\end{subequations}

\begin{eqnarray}
Y_{4,0}=\frac{2}{z^2}.
\end{eqnarray}
In the above equations,
\begin{eqnarray}
g_b=\frac{z_p\exp(-\beta E_b)}{(z_p-1)\exp(-\beta E_b)+1}
\end{eqnarray}
and
\begin{eqnarray}
g_s=\frac{z_p\exp(-\beta E_s)}{(z_p-1)\exp(-\beta E_s)+1}
\end{eqnarray}
are called the bending energy factors, and $N_{ie}$ $(i=2~\text{or}~3)$ is defined by half of the number of runs of $i$ consecutive bonds in a single chain, where one of the bonds links a sticker with a non-sticker, and hence, $N_{2e}=N_{3e}=1$ for linear chains.

\bibliography{refs}

\begin{thebibliography}{57}
\bibitem [{pap()}]{paper_note1}%
  \BibitemOpen
  \href@noop {} {}\bibinfo {note} {A rigorous definition for the LCT
  configurational entropy density appears in Ref.~\citenum{JCP_119_5730}.
  Recent computations~\cite{JCP_141_234903} indicate that the ordinary entropy
  density (evaluated as $s=-\partial f/\partial T |_{\phi}$) is nearly
  identical to the configurational entropy density $s_c$ derived in
  Ref.~\citenum{JCP_119_5730}, probably because the lattice model is
  essentially devoid of vibrational contributions. Therefore, calculations of
  the configurational entropy simply follow the definition of the ordinary
  entropy in the present paper.}\BibitemShut {Stop}%
\end{thebibliography}

\end{document}